\documentclass[fleqn,twoside]{article}
\usepackage{espcrc2}
\usepackage{epsfig}
\usepackage{amsmath}
\usepackage{amssymb}
\usepackage{amsbsy}
\usepackage{amsfonts}
\usepackage{graphics}
\usepackage{latexsym}
\usepackage{epsf}
\usepackage{graphicx}
\usepackage[figuresright]{rotating}

\usepackage{graphics}
\usepackage{latexsym}

\title{\begin{flushleft}
\vspace*{-2cm}
{\normalsize DESY 03-193}\\[-0.3em]
{\normalsize LU-ITP 2003/030}\\[-0.3em]
{\normalsize Edinburgh-2003/30}
\vspace*{0.15cm}
\end{flushleft}
Structure of the Nucleon\thanks{Talk presented by G. Schierholz at 
LHP2003, Cairns, Australia.}}

\author{M. G\"ockeler\address{Institut f\"ur Theoretische Physik, 
Universit\"at Leipzig, D-04109 Leipzig, Germany}\address{
Institut f\"ur Theoretische Physik, Universit\"at Regensburg, 
D-93040 Regensburg, Germany},
T.R. Hemmert\address{Physik-Department, Theoretische Physik, Technische
Universit\"at M\"unchen, 85747 Garching, Germany},
R. Horsley\address{School of Physics, University of Edinburgh, 
Edinburgh EH9 3JZ, UK},
D. Pleiter\address{John von Neumann-Institut f\"ur Computing 
NIC, Deutsches Elektronen-Synchrotron DESY,\\ 15738 Zeuthen, Germany},
P.E.L. Rakow\address{Theoretical Physics Division, Department of 
Mathematical Sciences, University of Liverpool,\\ Liverpool L69 3BX, UK},
A. Sch\"afer$^{\rm b}$,
G. Schierholz$^{\rm e}$\address{Deutsches Elektronen-Synchrotron  
DESY, 22603 Hamburg, Germany} and
W. Schroers\address{Center for Theoretical Physics, Massachusetts Institute 
of Technology, Cambridge, MA 02139, USA} \\[0.8em]
{QCDSF Collaboration}}

\begin{document}

\begin{abstract}
Generalized parton distributions provide information on the longitudinal and
transverse distribution of partons in the fast moving nucleon. Furthermore,
they contain information on the spin structure of the nucleon. First results
of a lattice study of generalized parton distributions are presented. 
\end{abstract}

\maketitle

\section{INTRODUCTION}

Understanding the internal structure of hadrons in terms of quarks and gluons
(partons), 
and in particular how quarks and gluons provide the binding and spin of the
nucleon, is one of the outstanding problems in particle physics. 

\begin{figure}
\begin{center}
\epsfig{file=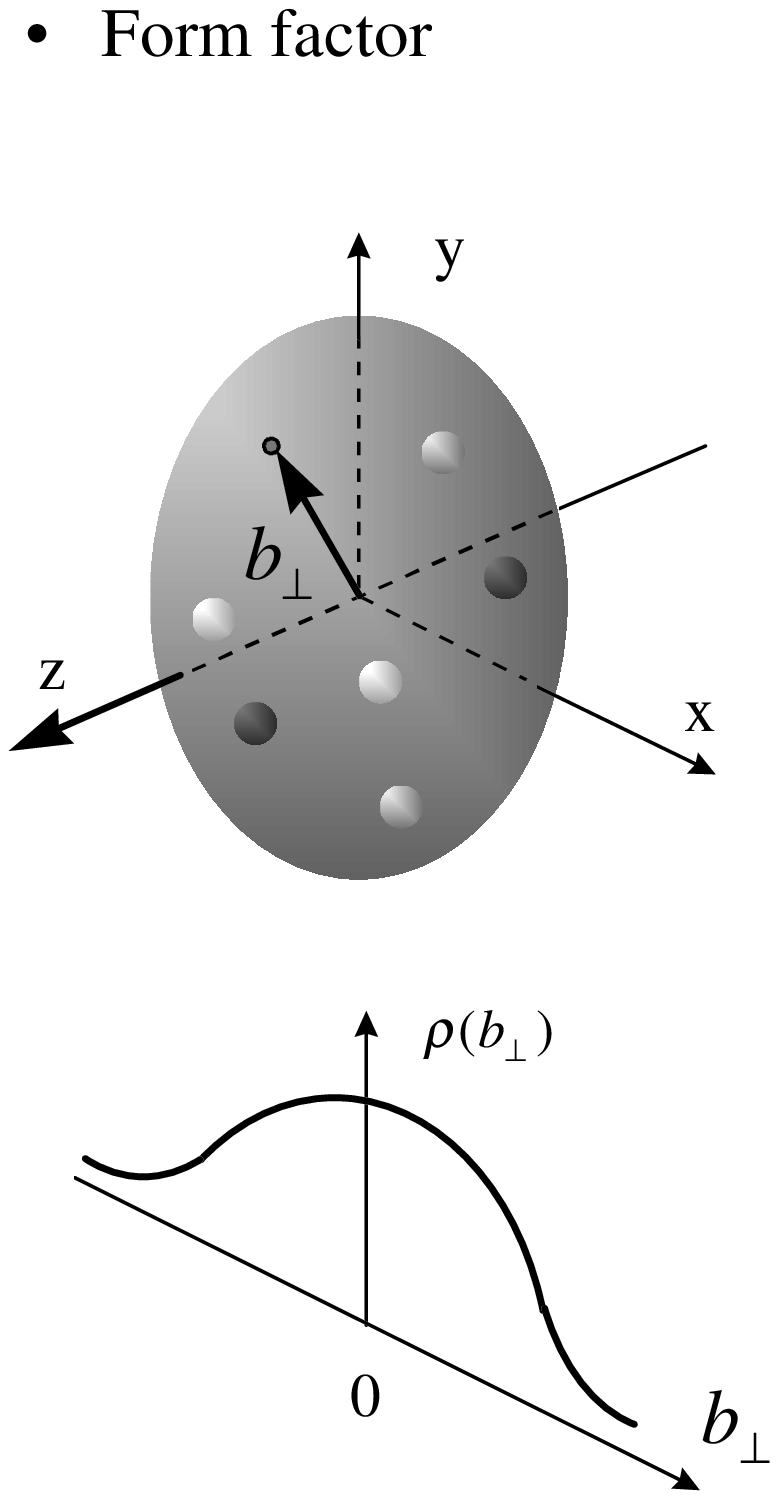,width=3.5cm,height=5.5cm,clip=}\\[0.5em] 
\hspace*{-0.5cm}
\epsfig{file=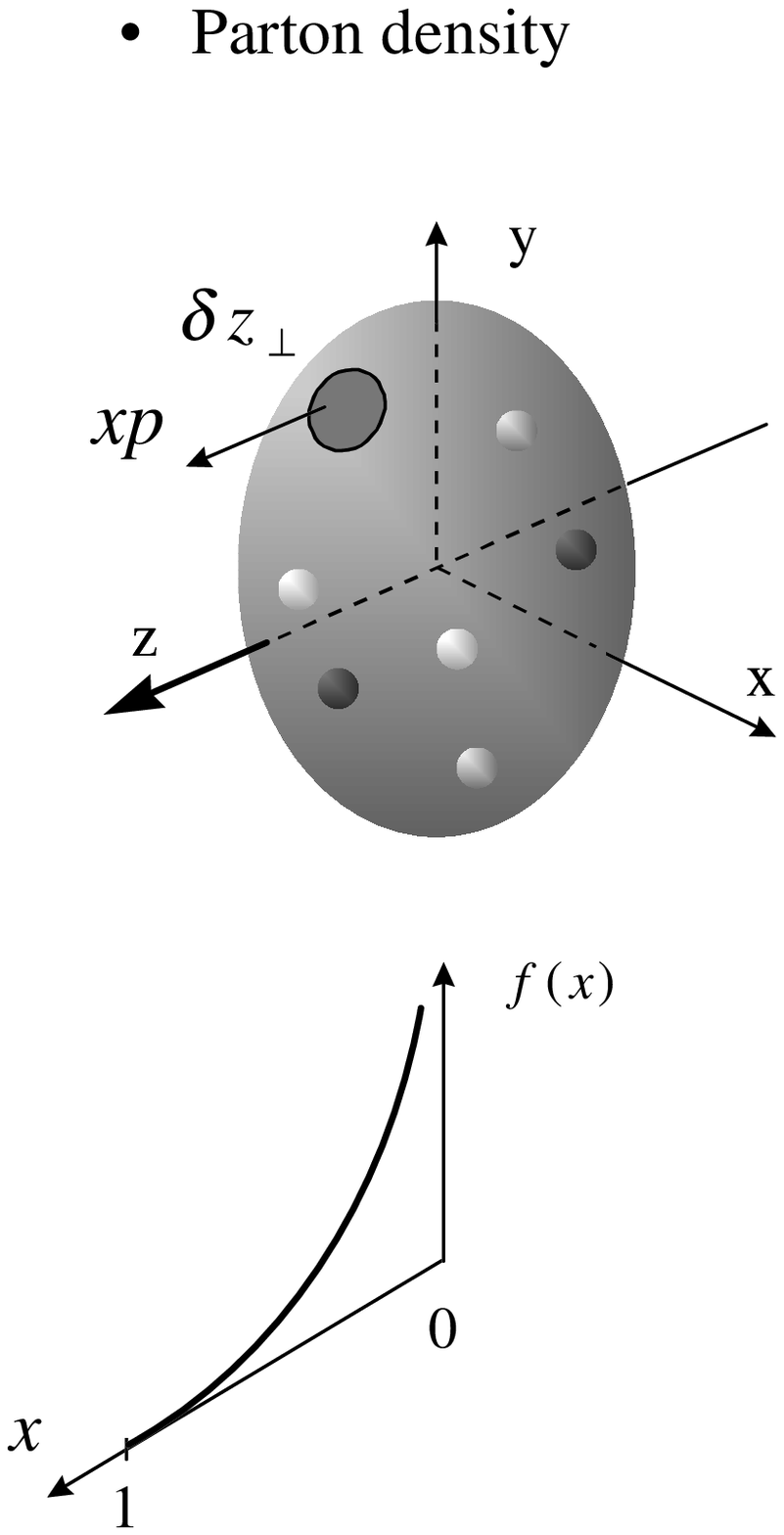,width=3.5cm,height=5.5cm,clip=}\\[0.5em] 
\hspace*{-1.25cm} \epsfig{file=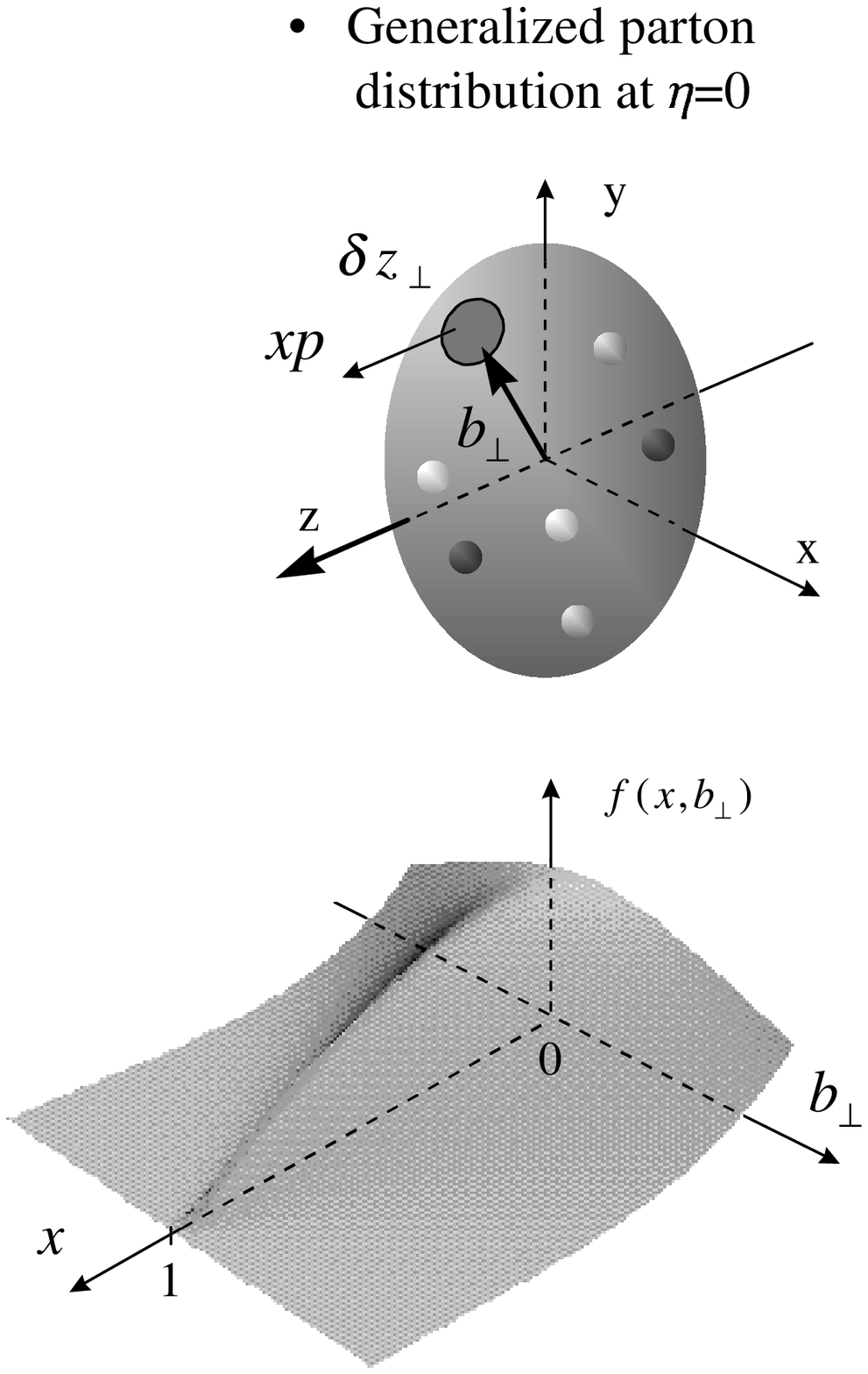,width=5cm,height=6cm,clip=}
\end{center}
\caption{Probabilistic interpretation of form factors, parton distributions
  and generalized parton distributions at $\eta=0$~\cite{BM} in the infinite
  momentum frame (resolution: $\delta z_\perp = 1/Q$).}
\end{figure}

Electromagnetic form factors provide information on the spatial distribution
of charge and magnetization in the nucleon, irrespective of the partons'
momenta and independent of the resolution scale. Parton distributions measure
the probability $|\psi(x)|^2$ of finding a parton with fractional longitudinal
momentum $x$ in the fast moving nucleon at a given transverse resolution
$1/Q$, while no information on the transverse distribution of partons is
provided. Generalized parton distributions (GPDs)~\cite{GPD} describe the
coherence of two 
different hadron wave functions $\psi^\dagger(x+\xi/2)\,\psi(x-\xi/2)$, one 
where the parton carries fractional momentum $x+\xi/2$ and one where this
fraction is $x-\xi/2$, from which further information on the transverse
distribution of partons can be drawn~\cite{Diehl}. In the limit where the
momentum transfer $\Delta$ on the nucleon is purely transverse, 
i.e. $\Delta=(0,\mathbf{\Delta}_\perp)$ and $\xi=\eta=0$, GPDs regain a
probabilistic interpretation~\cite{Bu}. When Fourier transformed to  
impact parameter space, they measure, for example, the probability of finding
a parton of 
momentum fraction $x$ at the impact parameter $\mathbf{b}_\perp$. In Fig.~1 we
compare the information provided by form factors, parton distributions and
GPDs. In other words, GPDs extend the well-known Feynman parton distribution
and  
electromagnetic form factors of the nucleon to new kinematic dimensions. In
the forward limit ($\Delta = 0$), these distributions reduce to the Feynman
parton distributions. 

Moments of GPDs are amenable to lattice calculations~\cite{QCDSF-1,MIT}.
Thus, they offer promising ways to link phenomenological observations to 
first principle theoretical considerations. In this talk we will report on
recent quenched results obtained by the QCDSF 
collaboration~\cite{QCDSF-1}. The calculation employs non-perturbatively
improved Wilson fermions and non-perturbative renormalization
factors~\cite{NP} of the operators whose matrix elements are involved. Generally, improved Wilson
fermions show only a very mild 
cut-off dependence~\cite{QCDSF}. We may therefore restrict ourselves to
calculations done on the $16^3\,32$ lattice at $\beta=6.0$.

\begin{figure}[!ht]
\vspace*{-0.1cm}
\begin{center}
\epsfig{file=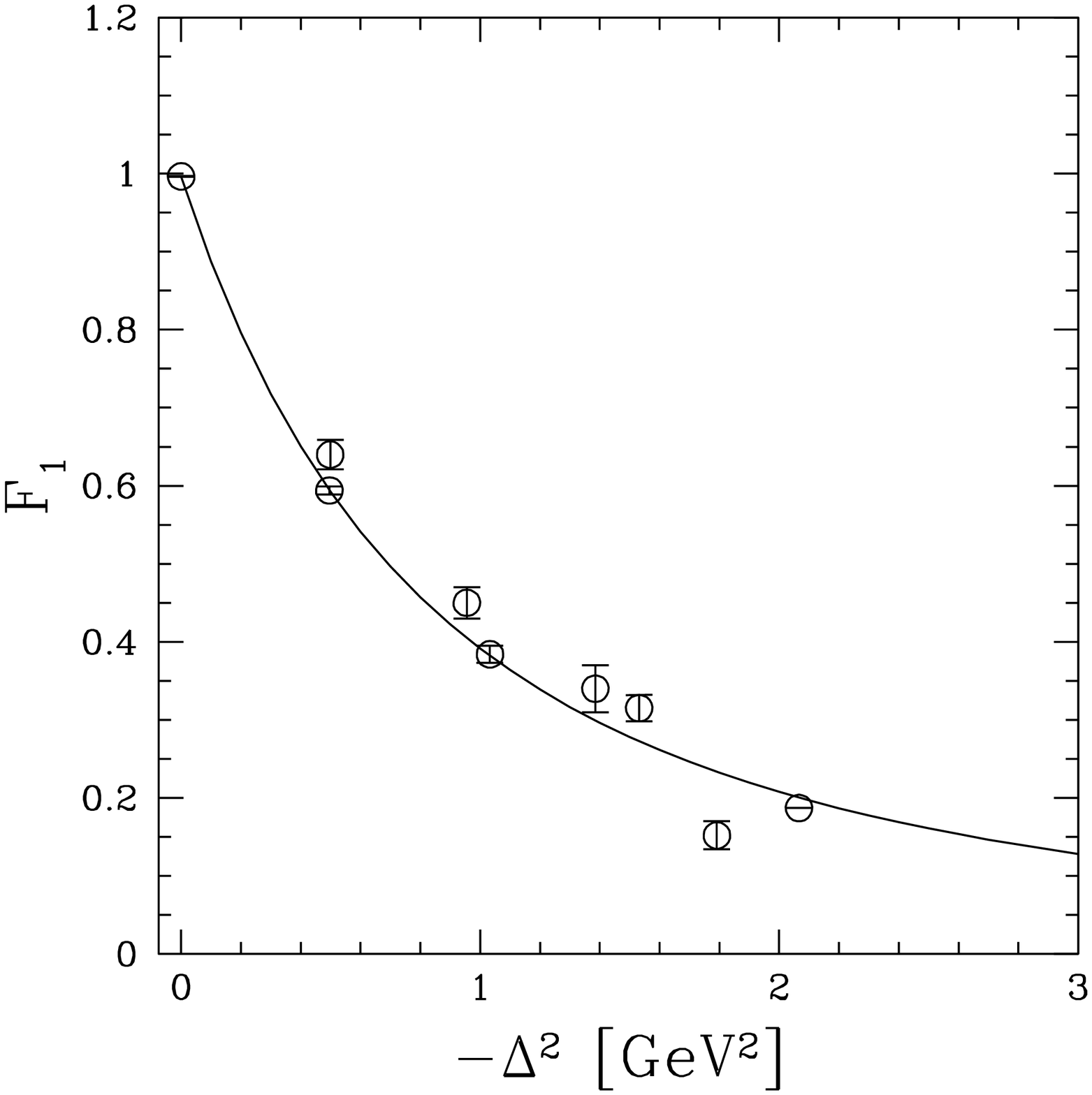,width=7.5cm,clip=}
\end{center}
\vspace*{-0.5cm}
\caption{The Dirac form factor $F_1(\Delta^2)$ of the proton at $m_\pi =
  450$ MeV, together with a dipole fit.} 
%
\vspace*{0.75cm}
\begin{center}
\epsfig{file=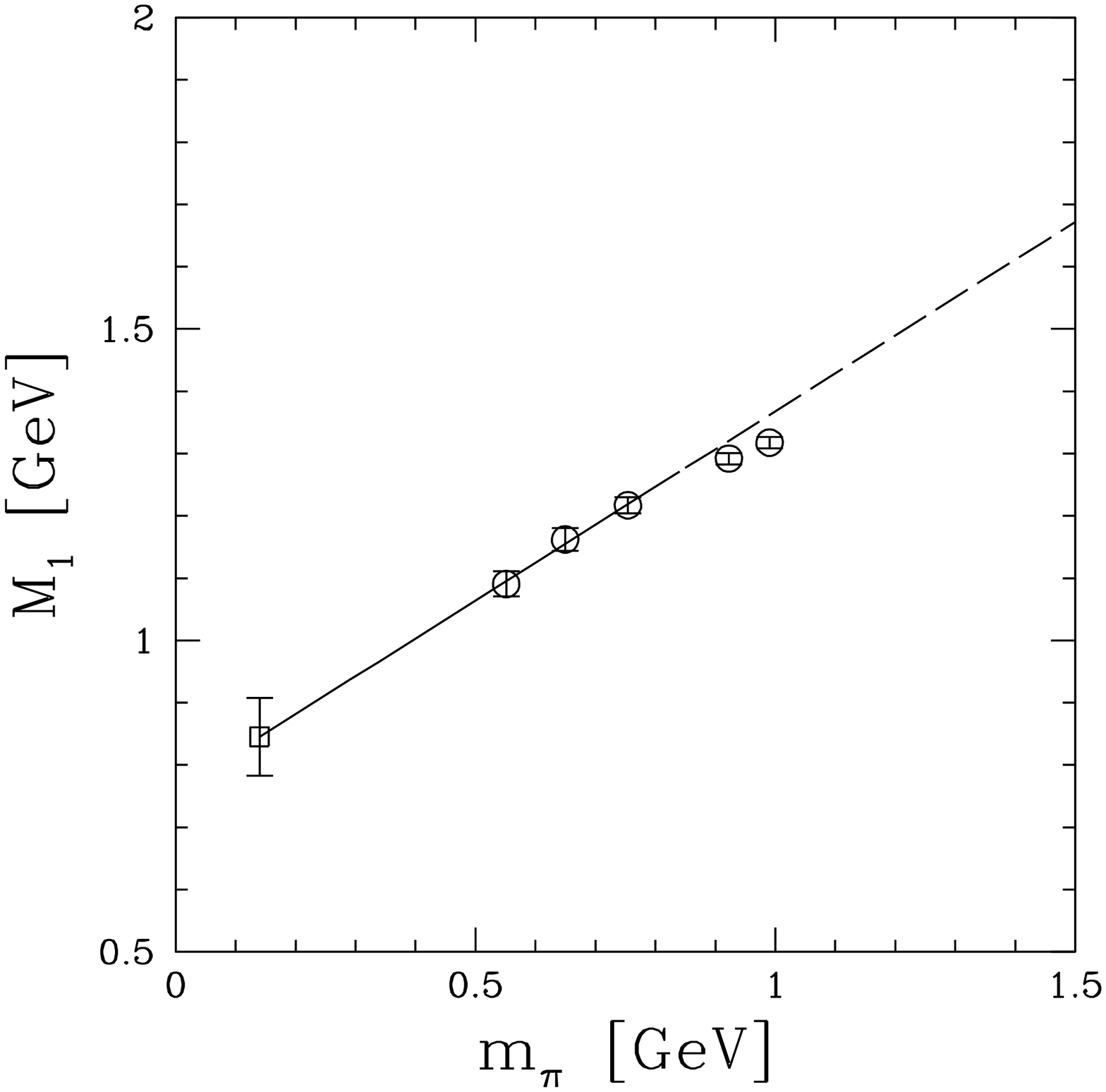,width=7.5cm,clip=}
\end{center}
\vspace*{-0.5cm}
\caption{The dipole mass $M_1$ as a function of $m_\pi$, together with a
  linear extrapolation to the physical pion mass ($\Box$).}
\vspace*{-1.5cm}
\end{figure}

\section{(GENERALIZED) FORM FACTORS}

We shall consider the GPDs $H_q$ and $E_q$ in the nucleon, where $q = u, d,
\cdots$ denotes the flavor of the struck quark. We will not 
consider the gluon sector here. 

The zeroth moments of $H_q$ and $E_q$
give the electromagnetic Dirac and Pauli form factors:  
\begin{equation}
\begin{split}
  \int_{-1}^1 \mbox{d}x\, H_q(x,\xi,\Delta^2) &= F^q_1(\Delta^2) \,, \\
  \int_{-1}^1 \mbox{d}x\, E_q(x,\xi,\Delta^2) &= F^q_2(\Delta^2) \,.
\end{split}
\end{equation}
Both form factors have been computed on the lattice~\cite{QCDSF,Thomas}. In 
Fig.~2 we show a typical result for the Dirac form factor of the proton, $F_1 =
F_1^u + F_1^d$. The solid curve is a dipole fit of the form
\begin{equation}
F_1(\Delta^2) = F_1(0)/(1-\Delta^2/M_1^2)^2\,.
\label{df}
\end{equation}
In Fig.~3 the dipole mass $M_1$ is extrapolated to the physical pion mass. We
find good agreement with the phenomenological value, which is
approximately equal to the physical $\rho$ and $\omega$ mass.

The first moment gives us generalized form factors of a class of tensor
operators:
\begin{equation}
\begin{split}
  \int_{-1}^1 \mbox{d}x\, x\, H_q(x,\xi,\Delta^2) &= A_2^q(\Delta^2) + 
  \xi^2 C_2^q(\Delta^2)\,,\\
  \int_{-1}^1 \mbox{d}x\, x\, E_q(x,\xi,\Delta^2) &= B_2^q(\Delta^2) - 
  \xi^2 C_2^q(\Delta^2)\,,\label{m}
\end{split}
\end{equation}
\begin{figure}[!t]
\begin{centering}
\epsfig{figure=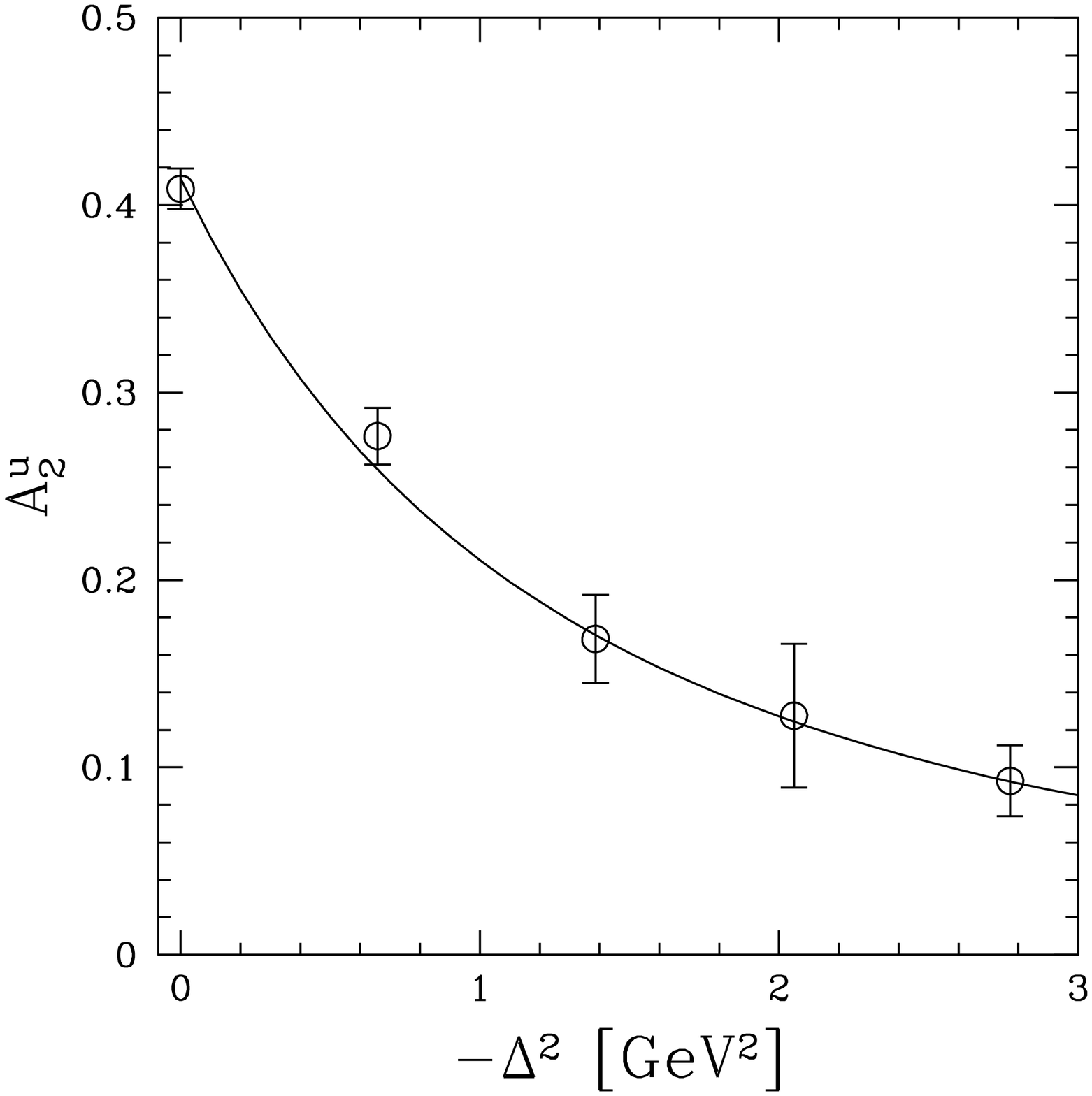,height=5.8cm,width=5.8cm}\\
\epsfig{figure=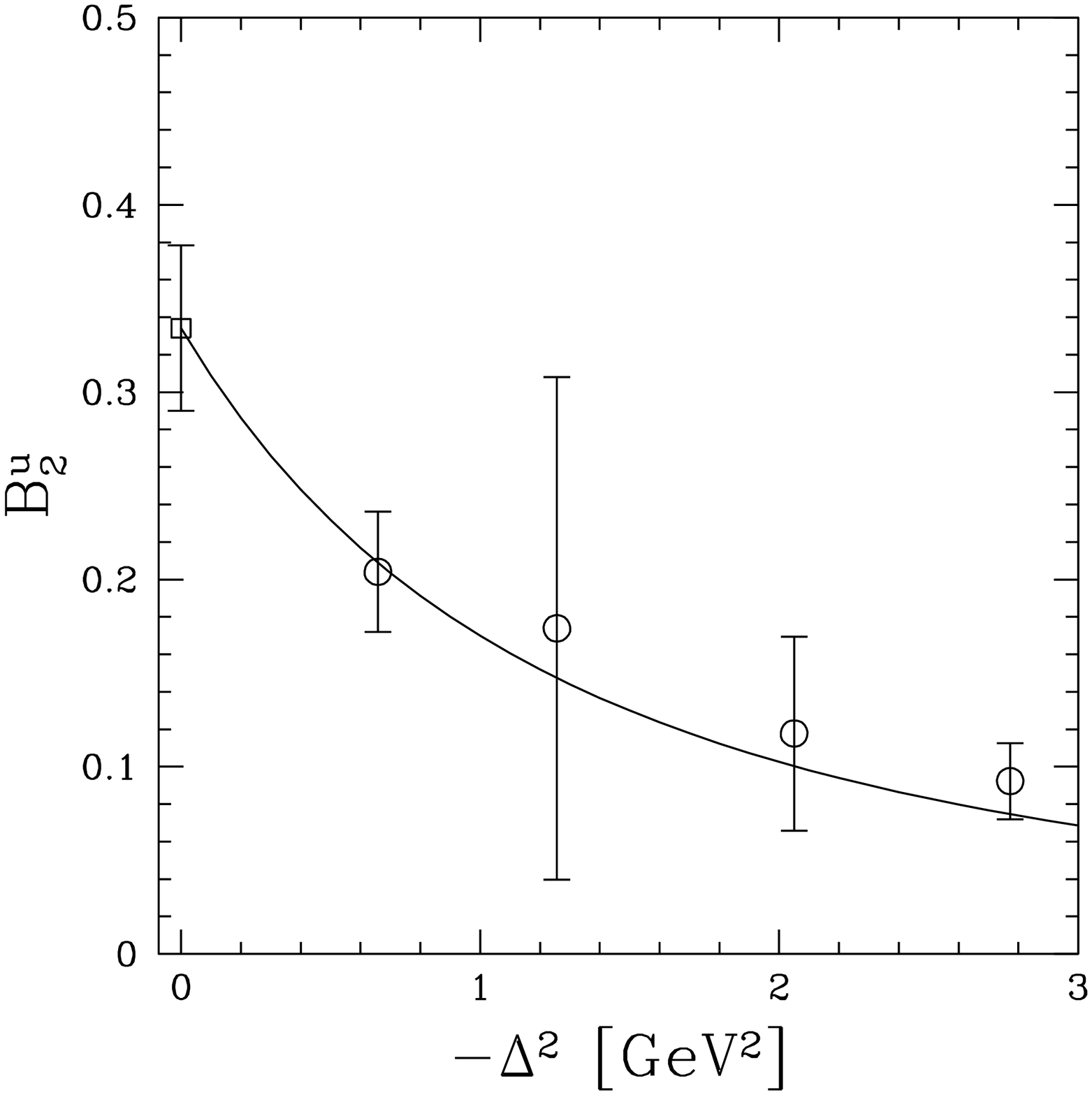,height=5.8cm,width=5.8cm}\\
\epsfig{figure=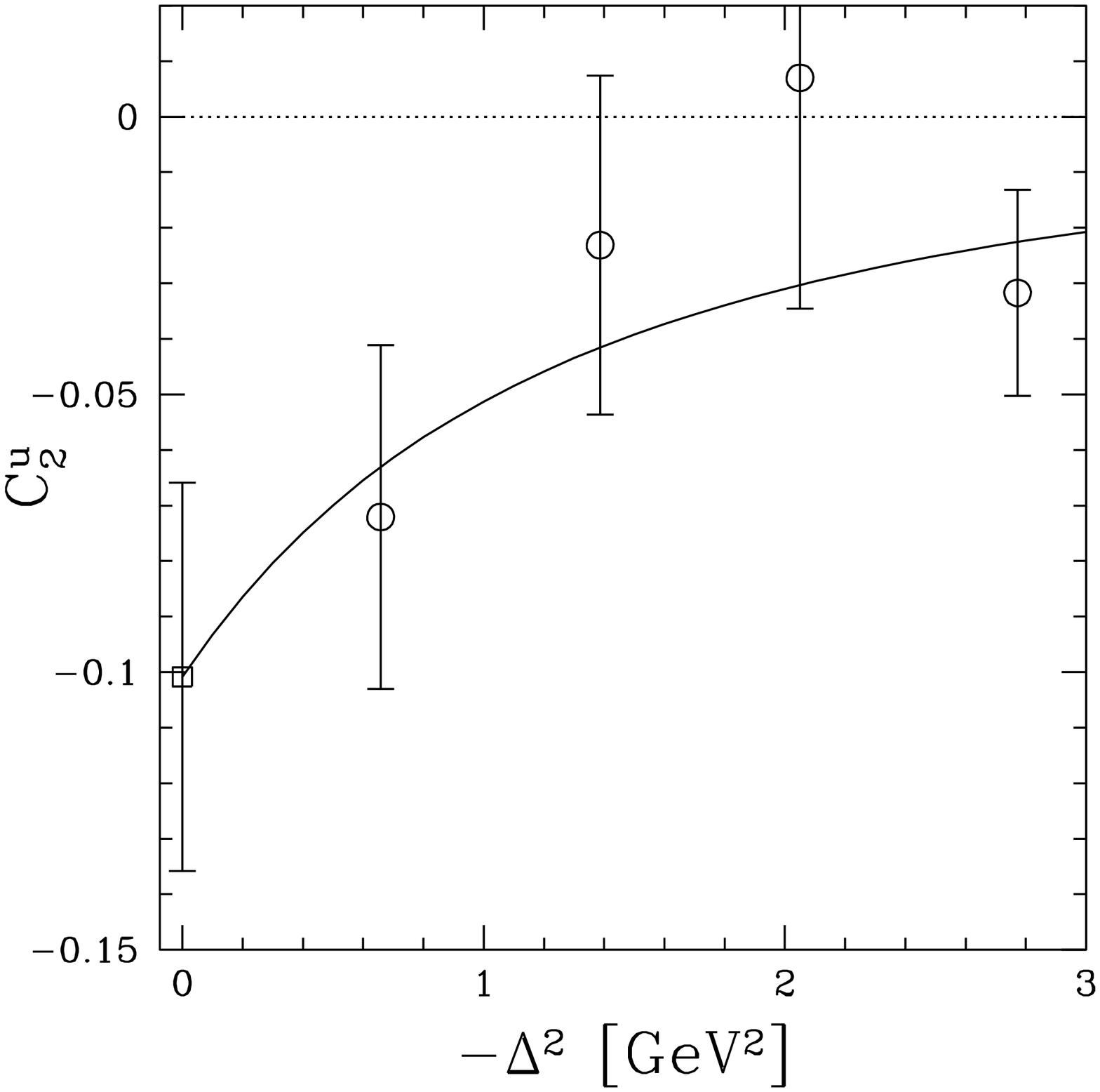,height=5.8cm,width=5.8cm}\\
\end{centering}
\caption{The generalized form factors $A_2^u$, $B_2^u$ and $C_2^u$ at 
$m_\pi=450$ MeV, together with the dipole fit and the extrapolated values 
at $\Delta^2=0$ ($\Box$).}
\vspace*{-1.25cm}
\end{figure}
\begin{figure}[!h]
\begin{centering}
\epsfig{figure=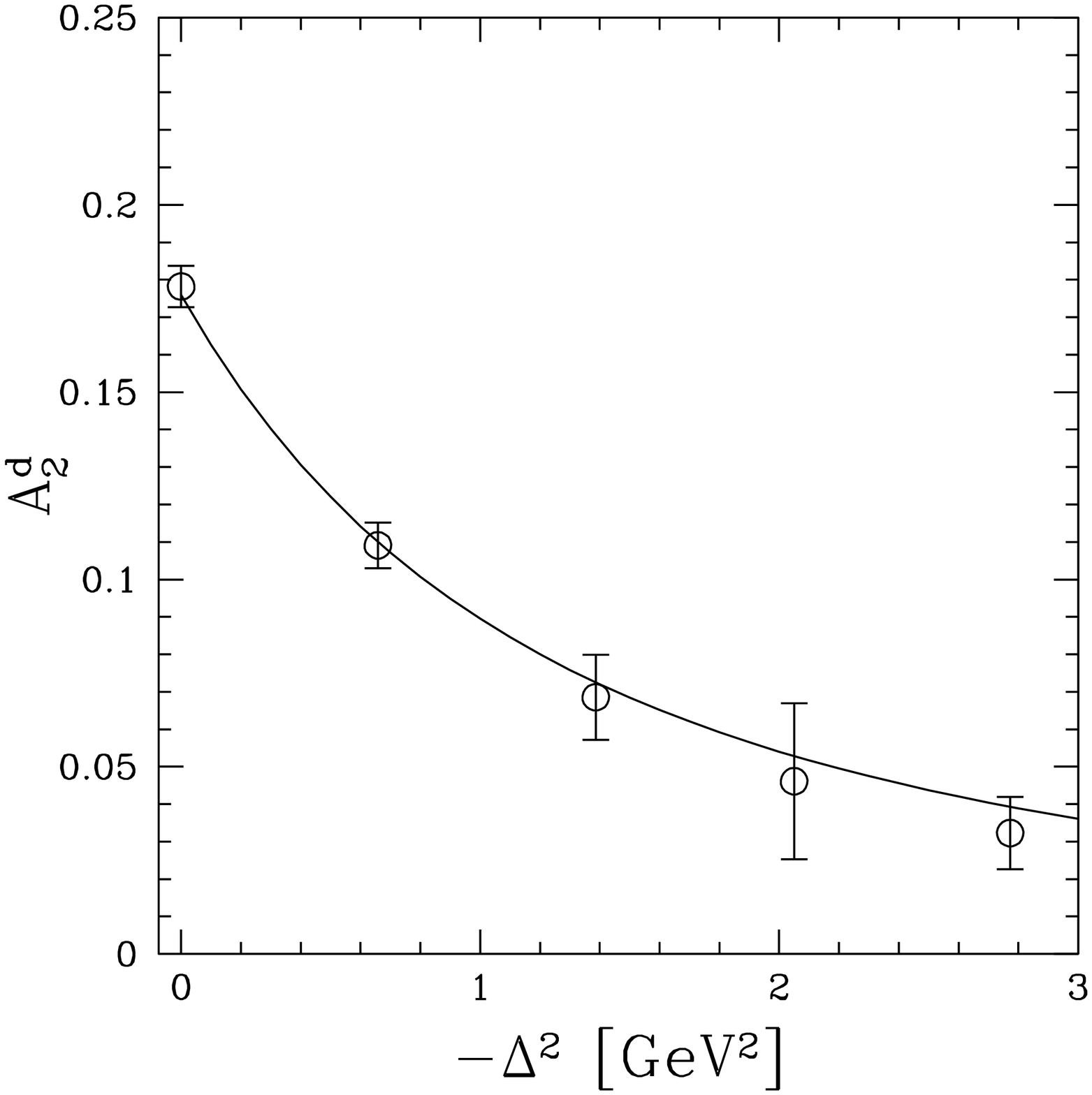,height=5.8cm,width=5.8cm}\\
\epsfig{figure=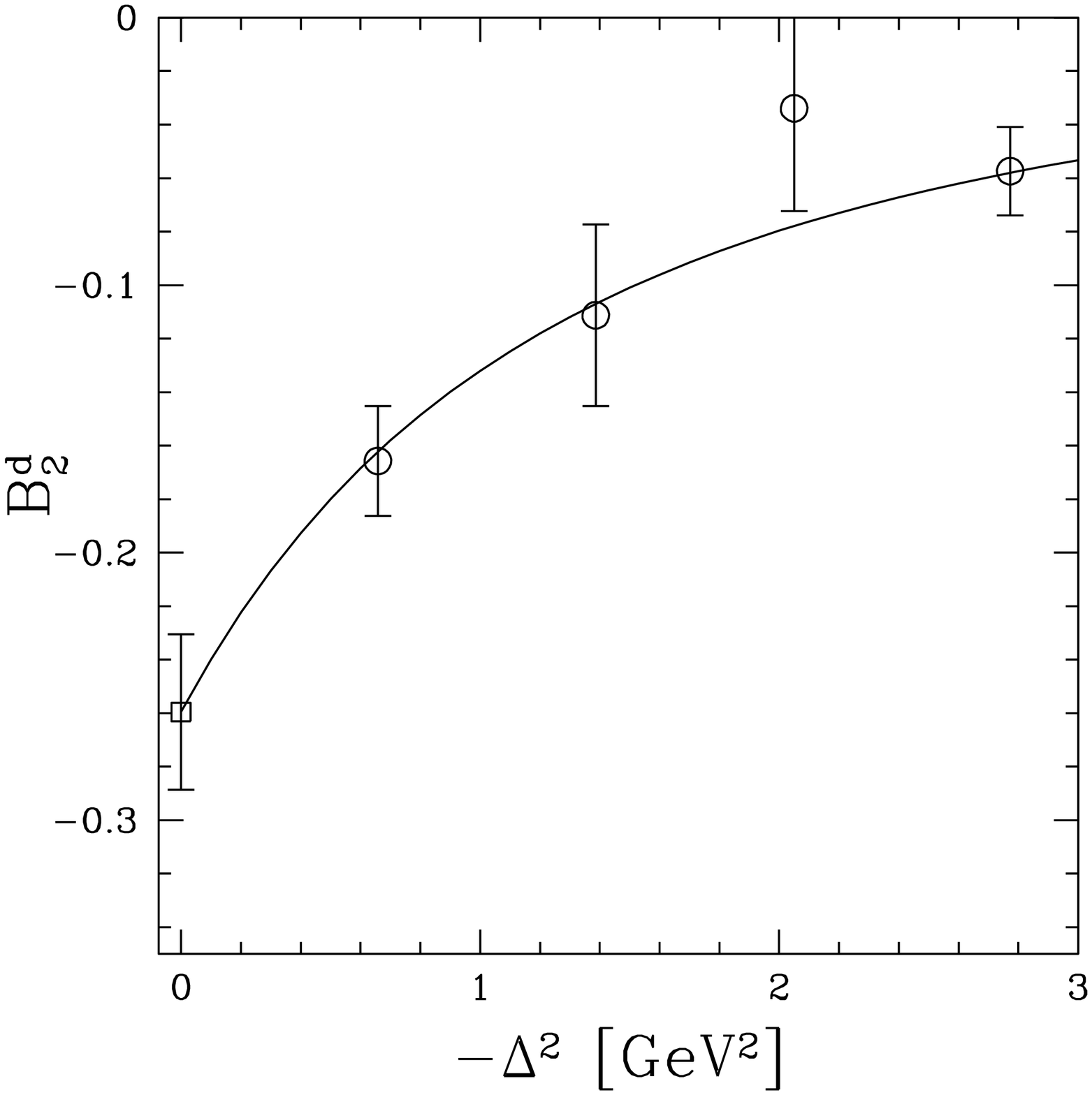,height=5.8cm,width=5.8cm}\\
\epsfig{figure=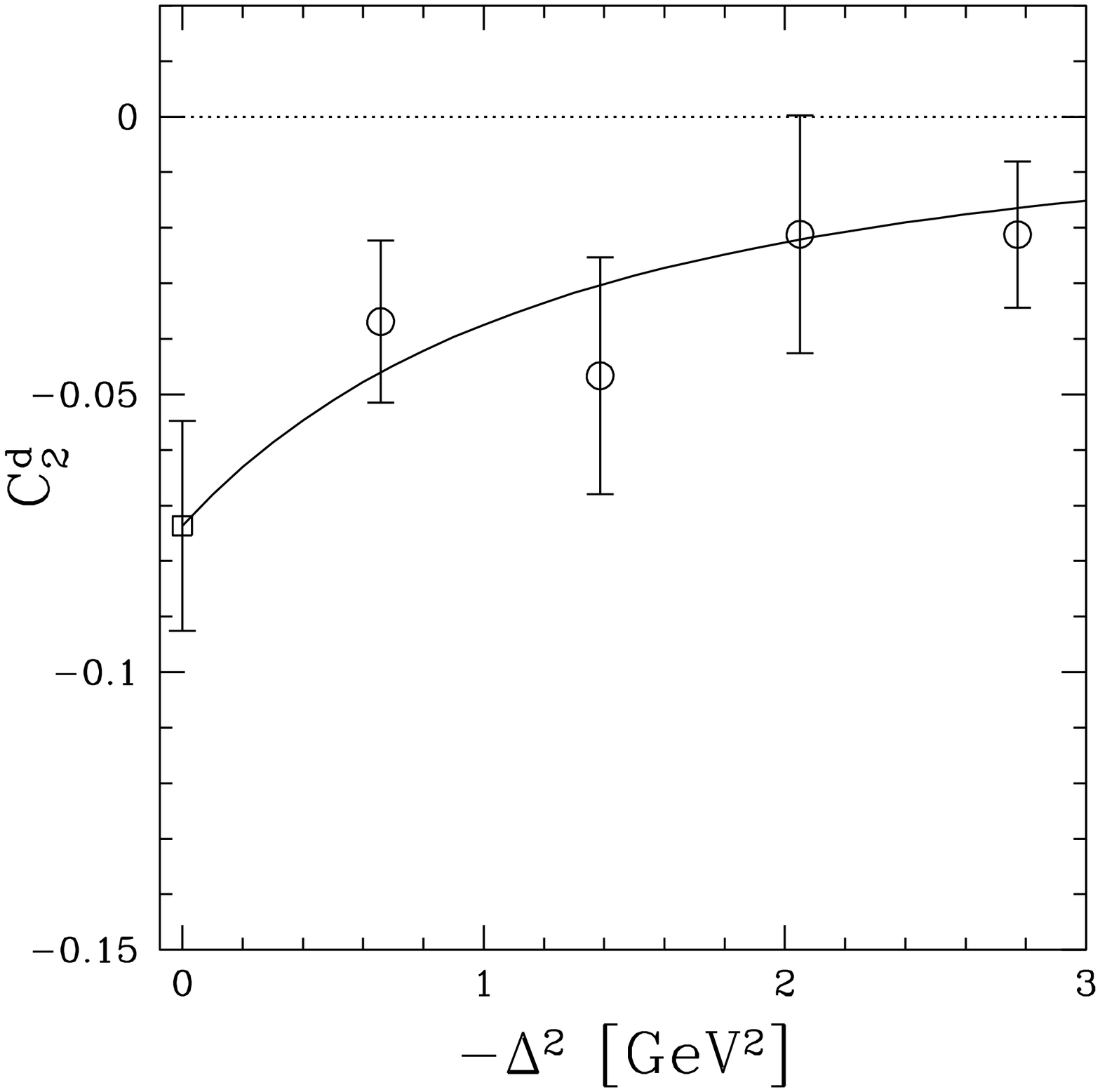,height=5.8cm,width=5.8cm}\\
\end{centering}
\caption{The generalized form factors $A_2^d$, $B_2^d$ and $C_2^d$ at 
$m_\pi=450$ MeV, together with the dipole fit and the extrapolated values 
at $\Delta^2=0$ ($\Box$).}
\vspace*{-1.25cm}
\end{figure}

\vspace*{-0.5cm}
\noindent where $A_2^q(\Delta^2)$, $B_2^q(\Delta^2)$ and 
$C_2^q(\Delta^2)$ are given by the nucleon 
matrix elements of the energy-momentum tensor (EMT):
\begin{equation}
\begin{split}
  \langle p'|{\cal O}^q_{\lbrace\mu\nu\rbrace}| p\rangle &\equiv
  \frac{\mbox{i}}{2} \langle p'|\bar{q}\gamma_{\lbrace\mu} 
  {\stackrel{\leftrightarrow}{D}}_{\nu\rbrace} q | p\rangle \\
   &= A_2^q(\Delta^2)\, \bar{u}(p')\gamma_{\lbrace\mu}\bar{p}_{\nu\rbrace} u(p)
  \\ 
   &- B_2^q(\Delta^2)\,\frac{\mbox{i}}{2m_N}\bar{u}(p')\Delta^\alpha
  \sigma_{\alpha\lbrace\mu}\bar{p}_{\nu\rbrace} u(p)  \\    
   &+ C_2^q(\Delta^2)\,\frac{1}{m_N}\bar{u}(p')u(p)\Delta_{\lbrace\mu}
  \Delta_{\nu\rbrace}\,.
\label{me}
\end{split}
\end{equation}
Here $m_N$ denotes the nucleon mass, $\bar{p}=\frac{1}{2}\left(p'+p\right)$, 
$\Delta=p'-p$, and curly brackets refer to symmetrization of indices and
subtraction of traces. The EMT has twist two and spin two. It is assumed to be
renormalized at the scale $Q$, which makes $A_2^q(\Delta^2)$, 
$B_2^q(\Delta^2)$ and $C_2^q(\Delta^2)$
scale and scheme dependent. For the classification of states of definite
$J^{PC}$ contributing to (\ref{me}) in the $t$-channel see \cite{Ji&Lebed}. 
The so-called skewedness parameter $\xi$ is 
defined by $\xi = -n\cdot\Delta$, where $n$ is a light-like vector with 
$n\cdot\bar{p}=1$, and bounded by $|\xi| \leq
2\sqrt{\Delta^2/(\Delta^2-4m_N^2)}$. 
In the forward limit, $\Delta^2 = \xi = 0$, the first moment of $H_q$
reduces to the first moment of the unpolarized parton distribution:
\begin{equation}
A_2^q(0) = \langle x_q\rangle 
\equiv \int_0^1 \mbox{d}x\, x\, \big(q_\uparrow(x)+q_\downarrow(x)\big)\,,
\end{equation}
where $q_{\uparrow(\downarrow)} (x)$ is the distribution with quark
spin parallel (antiparallel) to the spin of the nucleon. 
Furthermore, we have~\cite{Ji}
\begin{equation}
\frac{1}{2} \big(A_2^q(0)+B_2^q(0)\big) = J_q \,,
\label{J}
\end{equation}
where $J_q$ is the angular momentum of the $q$ quark, and
\begin{equation}
J = \sum_q J_q
\end{equation}
is the total angular momentum of the nucleon carried by the quarks. 

\begin{figure}[!b]
\begin{centering}
\epsfig{figure=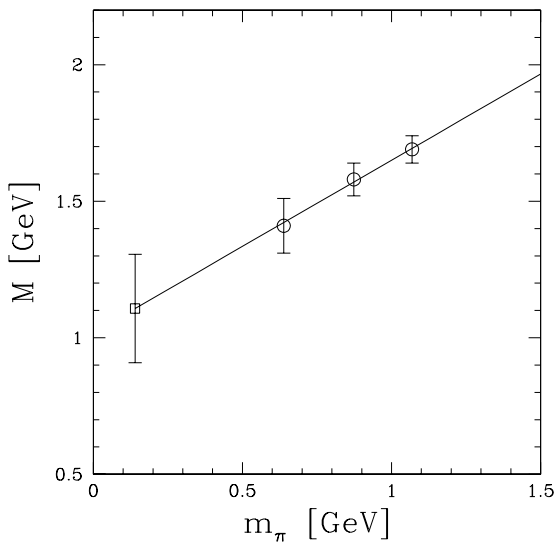,height=7.5cm,width=7.5cm}
\vspace*{-0.5cm}
\caption{The dipole mass $M$ as a function of $m_\pi$, together with a linear
extrapolation to the physical pion mass ($\Box$).} 
\end{centering}
\vspace*{-0.1cm}
\end{figure}

In Figs.~4 and 5 we show the generalized form factors 
$A_2^u(\Delta^2)$, $B_2^u(\Delta^2)$, $C_2^u(\Delta^2)$ and
$A_2^d(\Delta^2)$, $B_2^d(\Delta^2)$, $C_2^d(\Delta^2)$ of the proton, again
for $m_\pi=450$ MeV. As for the nucleon form factors, the generalized form
factors can be well described by the dipole ansatz
\begin{equation}
A_2^q(\Delta^2) = A_2^q(0)/(1-\Delta^2/M^2)^2 \,,
\label{dgpd}
\end{equation}
and similarly for $B_2^q$ and $C_2^q$. The numbers refer to the
$\overline{MS}$ scheme at the renormalization scale $Q = 2\,\mbox{GeV}$.
Fits of $A_2^u(\Delta^2)$ and $A_2^d(\Delta^2)$ 
give the same dipole mass $M$ within errors. We therefore fit
our data by a common dipole mass $M$, only depending on the quark mass. For a
reliable extrapolation to $\Delta^2 = 0$ it is important to cover a wide enough
range of $\Delta^2$. Our data do not favor a monopole behavior.

In Fig.~6 we show the dipole mass $M$ as a function of the pion mass. Again,
the numbers appear to lie on a straight line. A linear extrapolation in $m_\pi$
to the physical pion mass gives $M = 1.1(2)\,\mbox{GeV}$. This value is close 
to the physical masses of the $f_2$ and $a_2$ mesons, which supports the 
hypothesis of tensor meson dominance. The numbers $A_2^q(0)$, $B_2^q(0)$ and
$C_2^q(0)$ show little variation with the  quark mass and are extrapolated
quadratically in $m_\pi$ to the physical pion mass. We obtain
\begin{equation*}
\begin{tabular}{c|c|c|c}
$q$ & $A_2^q(0)$ & $B_2^q(0)$ & $C_2^q(0)$ \\ \hline 
$u$ & 0.400(22) & \;0.334(113) & -0.134(81) \\
$d$ & 0.147(11) & \;\!\!\!\!\!-0.232(77) & -0.071(42) \\
\end{tabular}
\end{equation*}

Higher moments of $H_q$ and $E_q$ may be obtained in a similar way. Knowing
sufficiently many moments, we can reconstruct $H_q(x,\xi,\Delta^2)$ and 
$E_q(x,\xi,\Delta^2)$ by inverse Mellin transform~\cite{DIS}.

\section{NUCLEON ANGULAR MOMENTUM}

The angular momentum decomposes,
in a gauge invariant way, into two contributions:
\begin{equation}
J_q = L_q + S_q \,,
\end{equation}
where $L_q$ is the orbital angular momentum and
\begin{equation}
S_q = \frac{1}{2} \Delta q \equiv \frac{1}{2} \int_0^1 \mbox{d}x\,
\big(q_\uparrow(x) - q_\downarrow(x)\big)
\end{equation}
is the spin of the quark. 
We know $\Delta q$ from separate calculations~\cite{QCDSF3,Deltaq}, 
so that $L_q$ can be computed from (\ref{J}).

\begin{figure}[t!]
\begin{centering}
\epsfig{figure=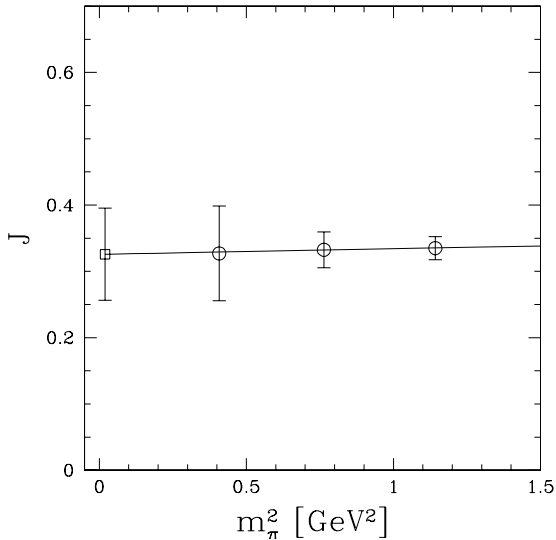,height=7.5cm,width=7.5cm}
\caption{The total angular momentum $J$, together with a
quadratic extrapolation to the physical pion mass ($\Box$).} 
\end{centering}
\vspace*{-0.35cm}
\end{figure}

In Fig.~7 we show the total angular momentum $J=J_u+J_d$. The dependence on 
the pion mass is
rather flat, as expected~\cite{Chen}. The errors are due to the relatively 
large statistical errors of 
$B_2^u$ and $B_2^d$ and the fact that $B_2^u$ and $B_2^d$ cancel to a large 
extent. 
A quadratic extrapolation in $m_\pi$ to
the physical pion mass gives 
\begin{equation*}
\begin{tabular}{c|c|c|c|c}
$J$ & $J_u$ & $J_d$ & $S_u$ & $S_d$ \\
\hline
\!0.33(7)\! & 0.37(6)\! & -0.04(4) \!& 0.42(1) \!& -0.12(1)\!\! \\
\end{tabular}
\end{equation*}
The numbers for $S_q$ 
refer to our latest results~\cite{QCDSF4}, computed from the non-perturbatively
improved axial vector current with non-perturbative renormalization 
factors. It turns out that the total angular momentum $J$ carried by the 
quarks amounts to 
$\approx 70\%$ of the spin of the (quenched) proton, leaving a contribution
of $\approx 30\%$ for the gluons. 
The major contribution is given by the $u$ quark, while the contribution of 
the $d$ quark is found to be negligible, which hints at strong pairing 
effects. 

From $J$ and $S$ we can compute $L_q$ now. The total 
orbital angular momentum of the (valence) quarks turns out to be consistent 
with zero:
\begin{equation}
L \equiv L_u + L_d = 0.03(7).  
\end{equation}
This indicates that (at virtuality $Q = 2\,\mbox{GeV}$) the parton's 
transverse momentum in the (quenched) proton is small. 

\section{GENERALIZED PARTON DISTRIBUTION}

If the dipole behavior (\ref{df}), (\ref{dgpd}) continues to hold for the 
higher moments as well, and if we assume that the dipole masses 
continue to grow in a Regge-like fashion, we may write
\begin{equation} 
\int_{-1}^1 \mbox{d}x\, x^n \,H_q(x,0,\Delta^2) = 
\langle x_q^n\rangle /(1-\Delta^2/M_{n+1}^2)^2 \,,
\label{H}
\end{equation}
with 
\begin{equation}
M_l^2 = \mbox{const.} + l/\alpha'\,,
\end{equation}
where $\mbox{const.} \approx -0.5$ GeV$^2$ and $1/\alpha' \approx 1.1$
GeV$^2$. We know $\langle x_q^n\rangle$ from previous lattice calculations for
$0 \leq n \leq 3$~\cite{QCDSF5}. That is sufficient to compute
$H_q(x,0,\Delta^2)$ by means of an inverse Mellin transform~\cite{DIS}.

\begin{figure*}[!t]
\begin{center}
\hspace*{-0.75cm}  \epsfig{file=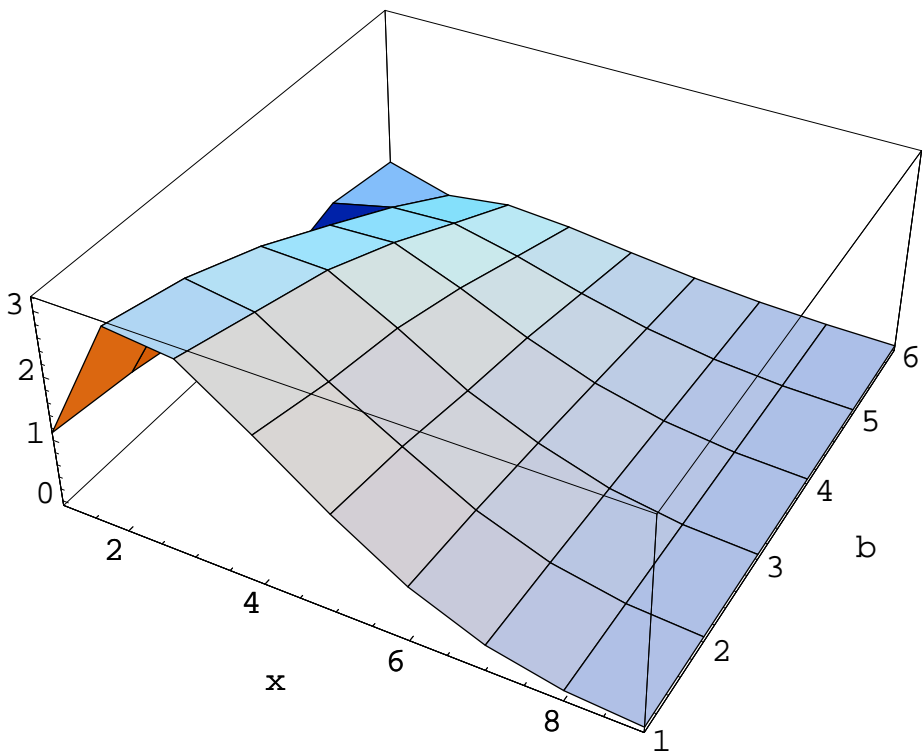,width=13cm}
\end{center}
\vspace*{-0.5cm}
\caption{The generalized parton distribution $xH_u(x,0,\mathbf{b}_\perp^2)$ as
  a function of $x$ and the impact parameter $b=|\mathbf{b}_\perp|$. The plot
  extends over the range $0.1 \leq x \leq 0.9$ and $0.1$ fm $\leq b \leq 0.6$
  fm.} 
\end{figure*}

Having done so, the desired probability distribution of finding a parton of
momentum fraction 
$x$ at the impact parameter $\mathbf{b}_\perp$ (see Fig.~1) is then obtained
by the Fourier transform of $H_q(x,0,\Delta^2)$:
\begin{equation}
H_q(x,0,\mathbf{b}_\perp^2)\! =\!\! \frac{1}{(2\pi)^2}\!\!\int\!\!
d^2\mathbf{\Delta}_\perp \, 
\mbox{e}^{\mbox{i}\mathbf{b}_\perp \mathbf{\Delta}_\perp\!} 
H_q(x,0,\mathbf{\Delta}_\perp^2)\,.
\end{equation}
In Fig.~8 we show a first attempt at a lattice calculation of
$H_q(x,0,\mathbf{b}_\perp^2)$ for the $u$ quark distribution in the proton.
The volume under the two-dimensional surface (taken over the full kinematical
range $0 \leq x \leq 1$ and $- \infty \leq b \leq \infty$) gives us back
$\langle x_u \rangle$. We see that at smaller values of $x$ the distribution in
$\mathbf{b}_\perp$ is rather broad (and different from the distribution
sketched in Fig.~1), 
while at larger values of $x$ it becomes somewhat more narrow. We also notice
that the peak of the distribution is shifted slightly towards larger values of
$x$ as $|\mathbf{b}_\perp|$ is increased.

\section{CONCLUSION}

We have presented a first quenched lattice estimate of the generalized parton
distribution 
$H_q(x,0,\mathbf{b}_\perp^2)$, assuming that the generalized form
factors of the higher operators (i.e. higher than the tensor operator) can be
described by 
a dipole ansatz as well, with dipole masses lying on a 
Regge trajectory. As a by-product we obtained the total and orbital angular
momentum of the nucleon carried by quarks. A calculation of the missing
generalized form factors is in progress.

\section*{ACKNOWLEDGEMENTS}

We like to thank the organizers of LHP2003, and in particular Alex Kalloniatis,
for a most productive and pleasant workshop, and hope for many more to come.
This work is supported in part by DFG and the EC under contract 
HPRN-CT-2000-00145. WS is supported by a Feodor-Lynen fellowship
and the U.S.~Department of Energy under cooperative research
agreement DF-FC02-94ER40818. The numerical calculations have been performed
on the APE100 at NIC (Zeuthen) and T3E at NIC (J\"{u}lich). We 
thank Dieter M\"uller for providing Figure~1.

\end{document}